\documentclass[aps,pra,twocolumn,superscriptaddress,floatfix]{revtex4-1} 
\usepackage[colorlinks,bookmarks=false,citecolor=blue,linkcolor=red,urlcolor=blue]{hyperref}
\usepackage{epsfig}
\usepackage{tikz}
\usepackage{graphicx,comment}

\usepackage{dcolumn}
\usepackage{bm}

\usepackage{amsmath,amssymb}

\usepackage{bbold}

\newcommand{\nn}{\nonumber}          
\newcommand{\bea}{\begin{eqnarray}}          
\newcommand{\eea}{\end{eqnarray}}

\allowdisplaybreaks

\begin{document}

\title{Spreading entanglement through pairwise exchange interactions}
\author{L. Theerthagiri}
\affiliation{Physics Division, School of Science and Technology, University of Camerino, I-62032 Camerino (MC),Italy}
\affiliation{Department of Physics, University of Naples Federico II,
I-80126 Napoli, Italy}
\author{R. Ganesh}
\email{r.ganesh@brocku.ca}
\affiliation{Department of Physics, Brock University, St. Catharines, Ontario L2S 3A1, Canada}

\date{\today}

\begin{abstract}
The spread of entanglement is a problem of great interest. It is particularly relevant to quantum state synthesis, where an initial direct-product state is sought to be converted into a highly entangled target state. In devices based on pairwise exchange interactions, such a process can be carried out and optimized in various ways. As a benchmark problem, we consider the task of spreading one excitation among $N$ two-level atoms or qubits. 
Starting from an initial state where one qubit is excited, we seek a target state where all qubits have the same excitation-amplitude -- a generalized-W state. 
This target is to be reached by suitably chosen pairwise exchange interactions. For example, we may have a a setup where any pair of qubits can be brought into proximity for a controllable period of time. 
We describe three protocols that accomplish this task, each with $N-1$ tightly-constrained steps. 
In the first, one atom acts as a flying qubit that sequentially interacts with all others. In the second, qubits interact pairwise in sequential order. In these two cases, the required interaction times follow a pattern with an elegant geometric interpretation. They correspond to angles within the spiral of Theodorus -- a construction known for more than two millennia. The third protocol follows a divide-and-conquer approach -- dividing equally between two qubits at each step. For large $N$, the flying-qubit protocol yields a total interaction time that scales as $\sqrt{N}$, while the sequential approach scales linearly with $ N$. For the divide-and-conquer approach, the time has a lower bound that scales as $\log N$. With any such protocol, we show that the phase differences in the final state cannot be independently controlled. 
For instance, a W-state (where all phases are equal) cannot be generated by pairwise exchange. 
\end{abstract}

\maketitle

\section{Introduction}
The synthesis of entangled states is an enduring problem in quantum science. This requires systematic protocols for transforming a direct-product state into a desired superposition of states. To mitigate decoherence effects, any such process
must be optimized to minimize operation time. This has inspired several studies on time-optimized protocols\cite{Burkard1999,Boozer2012,Geng2016,vanFrank2016,Bukov2018,Day2019,Tatsuhiko2022}. At the same time, it is important to ensure scalability. As quantum devices grow in qubit-number, entangling protocols must be able to operate within reasonable timeframes. 
This requires optimization with respect to qubit-number-complexity (operating time vs. number of qubits). Motivated by these ideas, we consider the simplest entanglement-spreading task -- that of spreading a single excitation equally among $N$ participating qubits. We impose a constraint informed by the design of multiple quantum architectures: this task is to be achieved solely by pairwise exchange interactions. We present three solutions and discuss their scaling with qubit number.

  The interest in entanglement spreading can be gauged from the large number of studies on the W state -- a prototypical entangled state where an excitation is equally spread over $N$ qubits\cite{Dur2000}. Many proposals have been put forward to synthesize the W-state\cite{Biswas2004,Perez2013} and many experiments have succeeded in creating it\cite{Haffner2005,Neeley2010,Grafe2014,Kagalwala2017}. The challenge in these protocols can be stated as follows: starting from an unentangled initial state with only one qubit excited, how can the excitation be spread equally among all qubits?
In this article, we take an approach that is inspired by \textit{mancala} games -- a family of games with a long history and wide geographical spread\cite{Russ1999}. They are played on a board with pits that contain pieces. In a typical game, a player picks pieces from one pit and distributes them over the other pits. Here, we have $N$ qubits that are analogous to $N$ pits. An excitation (an $\uparrow$ state or a 1-state) is initially stored in one qubit, analogous to pieces stored in a pit. The goal of the game is to spread the pieces evenly among $N$ pits. Below, we describe three protocols to achieve this goal and characterize their scaling with $N$.

We assume an architecture where qubits can undergo pairwise exchange interactions. Exchange interactions have been proposed as a mechanism for designing logic gates\cite{Loss1998}. They can be achieved in many settings. For example, with ultracold atoms in an optical lattice, a lattice-modulation can be used to induce an XXZ-exchange interaction\cite{Anderlini2007}. The time period and the strength of the interaction can both be controlled by tuning the modulation. In semiconductor qubits, exchange interactions can be induced in a similar fashion by tuning tunnelling barriers\cite{Petta2005,Maune2012,Yadav2021}. Alternatively, they can be mediated by a cavity-mode\cite{Woerkom2018} where the strength and duration can be controlled by varying the detuning.

\section{Entanglement by exchange} 
To set the stage, we begin by considering two two-level atoms (qubits), labelled A and B. They undergo an exchange interaction which entangles them. The degree of entanglement can be tuned by varying the interaction time, $t$.

We first discuss Heisenberg exchange as it leads to a simple form for the time-evolution operator. 
We then generalize to anisotropic exchange of the XXZ type. The results discussed in subsequent sections hold for any value of the XXZ anisotropy, including the Heisenberg limit. A Heisenberg exchange interaction between two qubits is described by the Hamiltonian 
\bea
\hat{H}_{AB} = J \Big[ 
\sigma_x^{A} \sigma_x^B + \sigma_y^{A} \sigma_y^B + \sigma_z^{A} \sigma_z^B 
\Big],
\eea
where $\sigma$'s are single-qubit operators encoded by Pauli matrices.
With the two qubits interacting for time $t$, the wavefunction undergoes unitary evolution. The time-evolution operator can be written in various forms. For our purposes, it is best written as
\bea
\hat{U}_{AB}(t) = e^{i t /2} \Big\{ \cos (t) ~\hat{\mathbb{1}}_{AB} -i  \sin(t) ~\hat{\Pi}_{AB} \Big\},~~~~
\label{eq.U}
\eea
where time $t$ is measured in units of $2\hbar/J$. The identity operator, $\hat{\mathbb{1}}_{AB}$, leaves both qubits unchanged. In contrast, $\hat{\Pi}_{AB}$ is the permutation operator that switches the states of A and B. In the $S_z$ basis ($\{ \uparrow\uparrow,\uparrow\downarrow,\downarrow\uparrow,\downarrow\downarrow\}$), it is given by 
\bea
\hat{\Pi}_{AB} = \left(\begin{array}{cccc}
1 & 0 & 0 & 0 \\
0 & 0 & 1 & 0 \\
0 & 1 & 0 & 0 \\
0 & 0 & 0 & 1 
\end{array}\right).
\eea
If one qubit is initially excited and the other is in the ground state, the permutation operator transfers the excitation from the former to the latter. As seen from Eq.~\ref{eq.U}, the amplitude for excitation transfer is $ \sin(t)$, while that for retaining the excitation at the same qubit is $\cos(t)$. By tuning the interaction time $t$, the `transferred weight' can be tuned. For a generic value of $t$, the final state is entangled with the excitation spread over two qubits.

We next consider a more general interaction Hamiltonian of the XXZ form,
\bea
\hat{H}_{AB}^\lambda = J \Big[ 
\sigma_x^{A} \sigma_x^B + \sigma_y^{A} \sigma_y^B + \lambda \sigma_z^{A} \sigma_z^B 
\Big],
\eea
where $\lambda$ is an anisotropy parameter. This Hamiltonian leads to the unitary time-evolution operator,
\bea
\nn \hat{U}_{AB}^\lambda (t) &=& e^{i\lambda t/2} \hat{P}_{\sigma_A \neq \sigma_B} \Big\{
\cos (t) ~\hat{\mathbb{1}}_{AB} -i  \sin(t) ~\hat{\Pi}_{AB} \Big\}\\
&+& e^{-i\lambda t/2} \hat{P}_{\sigma_A = \sigma_B},
\label{eq.Ulambda}
\eea
Here, $ \hat{P}_{\sigma_A = \sigma_B} $ is a projection operator onto the $\sigma_A = \sigma_B$ sector, where both qubits are in the same state. In this case, $\hat{U}_{AB}^\lambda (t) $ leaves the state unchanged (up to a global phase). In contrast, $ \hat{P}_{\sigma_A \neq \sigma_B} $ selects states where the qubits are in opposite states. Acting on such states, $\hat{U}_{AB}^\lambda (t)$ exchanges their states with probability amplitude $\sin(t)$. If the initial state had one qubit excited and one in the ground state, the final state will generically be entangled. The interaction time, $t$, controls the spread of the excitation across the two qubits.

Before stating the problem of interest, we note that the amplitudes in Eqs.~\ref{eq.U} and \ref{eq.Ulambda} are periodic in time with period $2\pi$.
In the following discussion, we always choose the shortest time that can effect a desired operation.

\section{Problem statement}
We consider $N$ two-level atoms (qubits). We assume a setup where pairs of atoms can be selected and made to interact for a specified period of time. For instance, this may involve bringing two atoms close to one other -- at a certain fixed distance and for a chosen time interval. Initially, the $N$ qubits are in a direct-product state represented as 
\bea
\vert \psi_{initial} \rangle = \vert \uparrow_1 \downarrow_2 \downarrow_3\ldots \downarrow_N \rangle.
\label{eq.initial}
\eea 
The first qubit is in the excited state while others are in the ground state. This can be viewed as one quantum of information stored in qubit-1. 

The target state is a generalized W-state given by
\bea
\vert \psi_{target} \rangle =\frac{1}{\sqrt{N}} \sum_{j=1}^N e^{i\phi_j}\vert \downarrow_1\ldots \downarrow_{j-1}\uparrow_j  \downarrow_{j+1}\ldots \downarrow_N \rangle.
\label{eq.target}
\eea
This is a sum of $N$ components, each having the excitation positioned at a different qubit. Each component has the same probability amplitude, but not necessarily the same phase. If all $\phi_j$'s were equal, this would be the well-known W-state\cite{Dur2000}. We do not place any restriction on $\phi_j$'s here. In fact, we will see below that $\phi_j$'s cannot be independently tuned.

In the following sections, we propose three protocols that take the initial state of Eq.~\ref{eq.initial} to the target state of Eq.~\ref{eq.target}. Our arguments hold for interactions of the XXZ type with any value for the anisotropy parameter, $\lambda$.

\section{Protocol with a single flying qubit}
\label{sec.flying}

\begin{figure}
\includegraphics[width=3.3in]{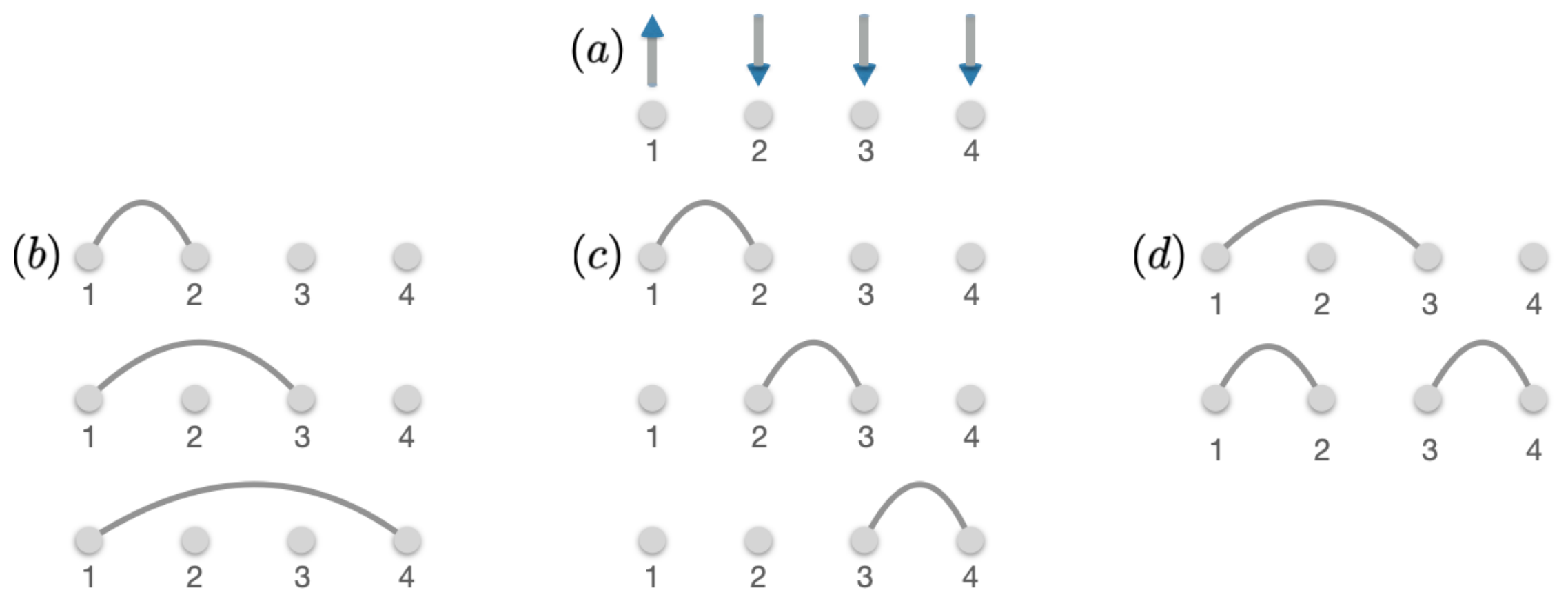}
\caption{Three protocols illustrated for a system with $N=4$ qubits. (a) The initial direct product state with one qubit excited and $N-1$ qubits in the ground state. (b) A flying-qubit protocol where qubit-1 interacts with each of the other qubits in order. (c) A sequential protocol where pairs of neighbouring qubits interact in succession. (d) A divide-and-conquer protocol where the system is arranged hierarchically in units of two qubits. At each stage, interactions act at one level of the hierarchy.}
\label{fig.protocols}
\end{figure}
We assume that one of the qubits can move freely and interact with each of the others. The qubit could be a photon or a vibration mode that can selectively couple to static qubits. In fact, the protocol discussed below was successfully used to generate a generalized W-state with trapped ions in 2005\cite{Haffner2005}. In this study, the role of the flying qubit was played by a vibration mode of a trapped-ion-chain. In the following discussion, we assume that this flying qubit is labelled as $j=1$. We further assume that this qubit is initially in the excited state while all other qubits ($j=2,3,\ldots,N$) are in the ground state. 

We propose a protocol where qubit-pairs interact in the following order: qubits $1$ and $2$ interact for time $t_{1,2}$, qubits $1$ and $3$ interact for time $t_{1,3}$, $\ldots$, qubits $1$ and $N$ interact for time $t_{1,N}$. Initially, qubits $1$ and $2$ begin in the state $\vert \uparrow_1 \downarrow_2 \rangle$. As they interact for time $t_{1,2}$, their state is acted upon by the time-evolution operator of Eq.~\ref{eq.U} or \ref{eq.Ulambda}. The resulting state is (up to a global phase) 
\begin{eqnarray}
\nonumber \cos (t_{1,2})\vert \uparrow_1 \downarrow_2 \downarrow_3 \ldots \downarrow_N\rangle - i \sin (t_{1,2})\vert \downarrow_1 \uparrow_2 \downarrow_3 \ldots \downarrow_N\rangle.
\end{eqnarray} 
After time $t_{1,2}$, qubit 2 does not interact with any of the other qubits. The component that is proportional to $\sin(t_{1,2})$ remains unchanged in amplitude, although it may accrue a phase. Therefore, in the final state, the probability amplitude for qubit 2 to be excited is $\sin (t_{1,2})$. Upon comparing with the target state of Eq.~\ref{eq.target}, we must have 
\begin{equation}
\sin (t_{1,2}) = \frac{1}{\sqrt{N}}.
\end{equation}
This fixes $t_{1,2}$. Subsequently, qubit 1 interacts with qubit 3 for time $t_{1,3}$. At the start of this process, the amplitude for qubit 1 to be excited is $\cos (t_{1,2}) = \sqrt{\frac{N-1}{N}}$. The probability amplitude for the excitation to be transferred to qubit 3 is given by a product of two amplitudes: (i) that for qubit 1 to be initially excited and (ii) that for the excitation to be transferred during the interaction. This is given by  
\begin{equation}
 \sqrt{\frac{N-1}{N}} \times \sin(t_{1,3}) = \frac{1}{\sqrt{N}}.
\end{equation}
We have set the amplitude to $1/\sqrt{N}$ in order to match the target state of Eq.~\ref{eq.target}. As qubit 3 does not interact after this step, it will always retain its amplitude through to the end. We obtain
\begin{equation}
\sin (t_{1,3}) = \frac{1}{\sqrt{N-1}}.
\end{equation}
At this point, the amplitude for qubit-1 to be excited is 
\begin{equation}
 \sqrt{\frac{N-1}{N}} \times \cos (t_{1,3}) =  \sqrt{\frac{N-1}{N}}  \times \sqrt{\frac{N-2}{N-1}} = \sqrt{\frac{N-2}{N}}.  
\end{equation}
At the next step, qubits 1 and 4 interact. The amplitude for an excitation to be transferred to qubit 4 is given by
\begin{equation}
\sqrt{\frac{N-2}{N}} \times \sin (t_{1,4}) = \frac{1}{\sqrt{N}}. 
\end{equation}
This fixes $\sin (t_{1,4}) = \frac{1}{\sqrt{N-2}}$. 
Proceeding in this manner, we find 
\begin{eqnarray}
\sin(t_{1,5}) = \frac{1}{\sqrt{N-3}},~\ldots,~
\sin (t_{1,N}) = \frac{1}{\sqrt{2}}.
\end{eqnarray}
These relations can be gathered into a general expression for the $j^\mathrm{th}$ time interval, 
\bea
t_{1,j+1} = \sin^{-1}\{1/\sqrt{N-j+1}\}.
\label{eq.timeprot1}
\eea 
Remarkably, these time periods have an elegant geometric interpretation. 
These are angles within the spiral of Theodorus, a geometric construction known since the 5th century BCE\cite{Davis1993,Gronau2004}. The spiral is constructed as a series of right-angled triangles. At each step, a unit line segment is drawn perpendicular to the hypotenuse of the previous step. This forms one side of a new right-angled triangle with a longer hypotenuse. This procedure leads to a sequence of points spiralling outwards. The n$^\mathrm{th}$ point is given as $(r_n,\theta_n)$ in polar coordinates. Here, $r_n =\sqrt{n}$ and $\theta_n$ is a monotonically increasing function of $n$. For large $n$, it is known\cite{Hlawka1980,Davis1993} that $\theta_n \sim 2\sqrt{n}$, with corrections that are subleading in powers of $n$.

\begin{figure}
\includegraphics[width=3.3in]{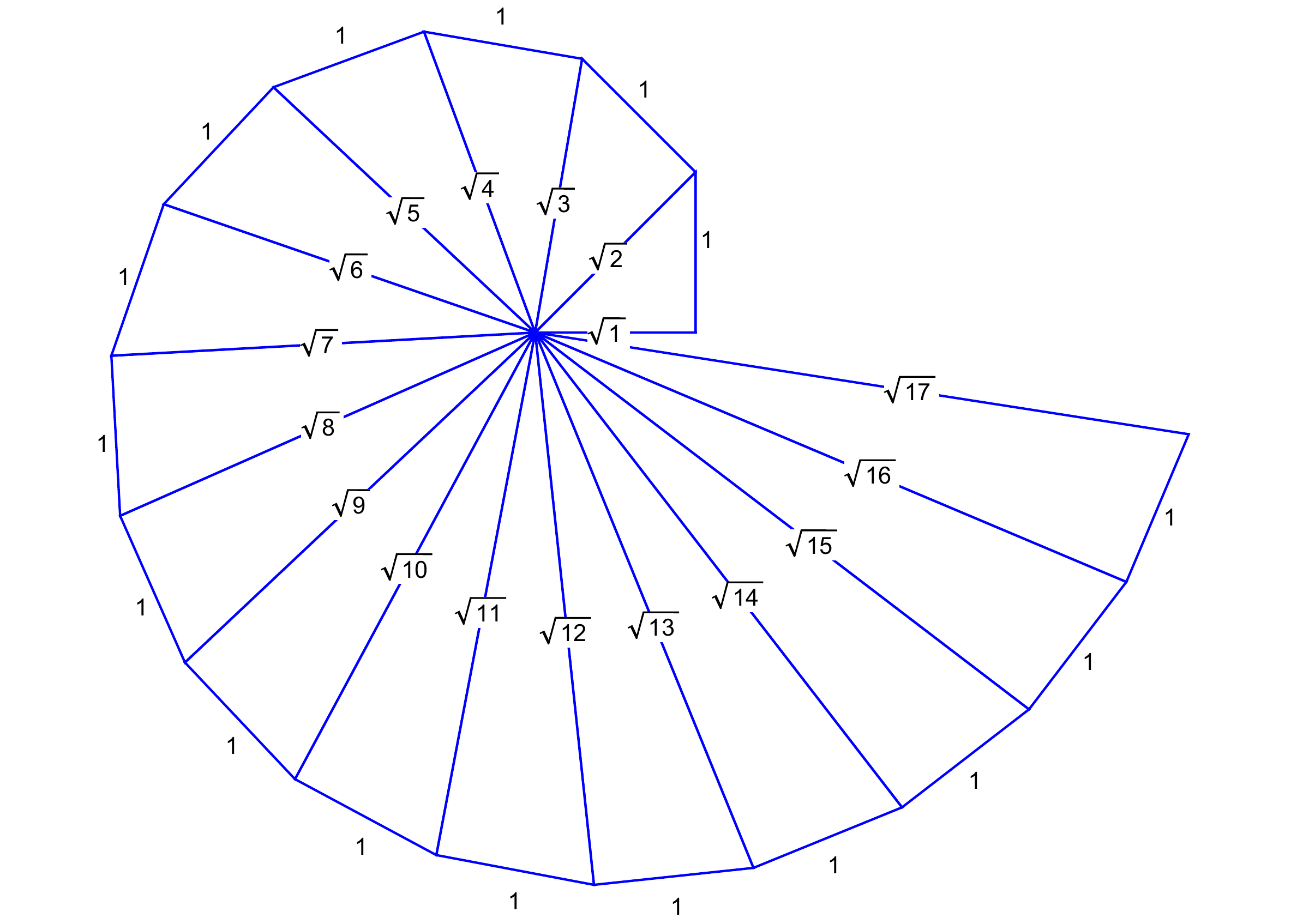}\\
\includegraphics[width=1.2in]{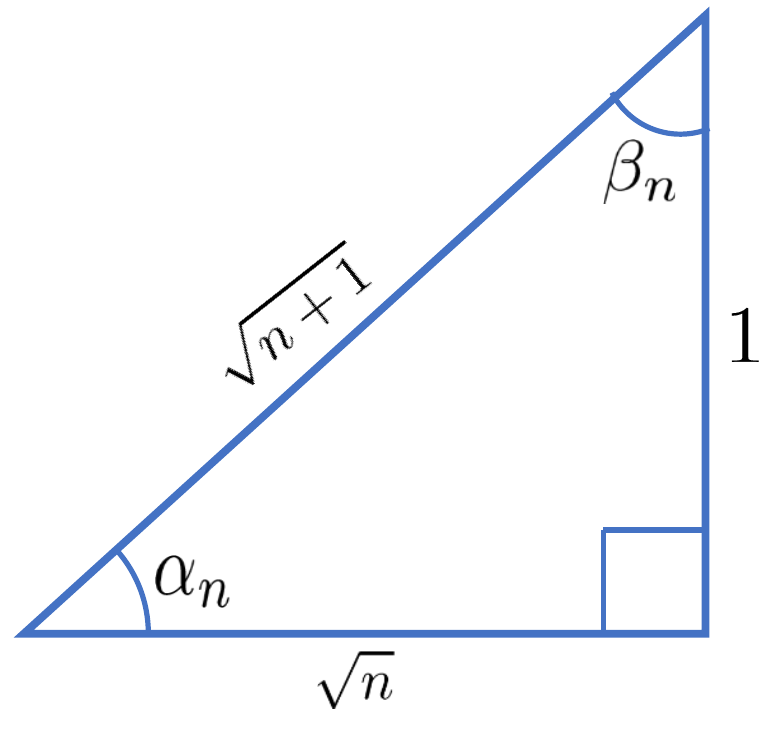}
\caption{Top: The spiral of Theodorus, constructed as a series of right-angled triangles. Bottom: The $n^\mathrm{th}$ triangle in the spiral, with sides $1$, $\sqrt{n}$ and $\sqrt{n+1}$. }
\label{fig.theo}
\end{figure}

Fig.~\ref{fig.theo} shows the angles as they appear in the spiral. The `interior angles', denoted as $\alpha_n$'s, are precisely the time intervals given in Eq.~\ref{eq.timeprot1},
\bea
\alpha_1 = t_{1,N};~~\alpha_2 = t_{1,N-1};~~\ldots;~~\alpha_{N-1} = t_{1,2}.
\eea
From the figure, it is clear that the interior angles decrease progressively, i.e., $\alpha_n$ monotonically decreases with $n$.  We deduce that the time intervals increase progressively, with $t_{1,2} < t_{1,3} < \ldots < t_{1,N}$. The total process time, excluding overheads such as rearranging qubits, is given by 
\begin{equation}
t_{flying} = t_{1,2} + t_{1,3} + \ldots + t_{1,N} = \sum_{j=1}^{N-1} \alpha_j  = \theta_N \approx 2 \sqrt{N}. 
\label{eq.tfly}
\end{equation} 
where $\alpha_j$'s are angles as shown in Fig.~\ref{fig.theo}. The sum over $\alpha_j$'s yields $\theta_N$, the angular coordinate of the $N^\mathrm{th}$ point of the Theodorus spiral. In the last step, we have used the approximate form for $\theta_N$ when $N$ is large. We arrive at the following result: this protocol yields a generalized W-state with the operation time scaling as $\sqrt{N}$ for large $N$.

\section{Protocol with sequential pairwise interactions}
\label{sec.sequential}
We next consider a protocol where qubit-pairs interact in the following order: qubits $1$ and $2$ interact for time $t_{1,2}$, qubits $2$ and $3$ interact for time $t_{2,3}$, $\ldots$, qubits $N-1$ and $N$ interact for time $t_{N-1,N}$. Initially, qubit $1$ is taken to be excited while all others are in the ground state. As qubits $1$ and $2$ interact, their state is acted upon by the time-evolution operator of Eq.~\ref{eq.U} or Eq.~\ref{eq.Ulambda}. A portion of the excitation can be transferred from qubit $1$ to $2$. At the next step, a portion of the excitation in qubit $2$ is transferred to $3$ and so on. 

Qubit $1$ is only modified during the first step. As a result, its final excitation-amplitude is determined at the first step alone. From Eq.~\ref{eq.U} or \ref{eq.Ulambda}, this is given by $\cos (t_{1,2})$ -- the amplitude for no excitation transfer occurring during the first step. In the final target state, the amplitude for qubit-$1$ to be excited must be $1/\sqrt{N}$, so that 
\begin{equation}
\cos(t_{1,2}) = \frac{1}{\sqrt{N}}.
\end{equation}
This fixes time $t_{1,2}$. After the first step, the amplitude for qubit-2 to be excited is given by $\sin(t_{1,2}) = \sqrt{\frac{N-1}{N}}$. During the second step, this excitation may be passed onto qubit-3. Beyond the second step, qubit-2 remains unchanged. As a result, the final amplitude for qubit-2 to be excited is given by $\sin(t_{1,2}) \times \cos(t_{2,3}) $. In the final target state, the amplitude for qubit-2 to be excited must be $1/\sqrt{N}$, so that
\begin{equation}
\sqrt{\frac{N-1}{N}} \cos(t_{2,3}) = \frac{1}{\sqrt{N}} \implies \cos (t_{2,3}) = \frac{1}{\sqrt{N-1}}.
\end{equation}
This fixes $t_{2,3}$. Considering each following step in the same fashion, we arrive at
\begin{eqnarray}
\cos(t_{3,4}) = \frac{1}{\sqrt{N-2}},
~~\ldots,~~
\cos(t_{N-1,N}) &=& \frac{1}{\sqrt{2}}.
\end{eqnarray}
These relations determine all time intervals in the problem, with $t_{j,j+1} = \cos^{-1}(1/\sqrt{N-j+1})$. These times are, once again, angles that appear in the spiral of Theodorus. As shown in Fig.~\ref{fig.theo}, they are `exterior angles' denoted as $\beta_n$'s. We have $\beta_1 = t_{N-1,N}$, $\beta_2 = t_{N-2,N-1}$, $\ldots$, $\beta_{N-1} = t_{1,2}$.

From Fig.~\ref{fig.theo}, we see that $\beta$'s increase monotonically with $n$. We conclude that the time intervals in this protocol are arranged in descending order: $t_{1,2} > t_{2,3} > \ldots > t_{N-1,N}$.

As seen from Fig.~\ref{fig.theo}, $\beta_j$ and $\alpha_j$ form a pair of complementary angles for any $j$. In Sec.~\ref{sec.flying}, the total operation time was written as a sum over $\alpha$-angles. Here, the total operating time is
\begin{eqnarray}
\nonumber t_{sequential} = t_{1,2} + t_{2,3} + \ldots + t_{N-1,N} = \sum_{j=1}^{N-1} \beta_j \\
= (N-1) \frac{\pi}{2} - \sum_{j=1}^{N-1} \alpha_j \approx (N-1) \frac{\pi}{2}- 2\sqrt{N}.~~~
\end{eqnarray}
We have used the result quoted in Eq.~\ref{eq.tfly} for the sum over $\alpha_j$. We conclude that the total operating time scales linearly with $N$ in this protocol. 

\section{Divide-and-conquer protocol}
\label{sec.divide}
The previous two sections present two protocols. 
In both, an initial excitation in one qubit is spread over $N$ qubits in serial fashion -- through a sequence of exchange interactions that must be executed in serial order. We now consider a third protocol that allows for parallel operations. At each step, we consider two qubits. One has a certain probability of being in the excited state, while the other is entirely in the ground state. An exchange interaction is carried out to equally spread the excitation-amplitude between the two qubits. 

This protocol is particularly suited for $N$'s that are powers of $2$, i.e., $N=2^M$ where $M$ is an integer. The protocol proceeds through $M$ stages where each stage may involve multiple pairwise interactions. For illustration, we take the example of 4 qubits ($M=2$). Initially, qubit-1 is excited while all others are in the ground state. During the first stage, qubit-1 and qubit-3 are made to interact. The interaction time is chosen such that the qubit-1 and 3 both acquire the same excitation-amplitude. That is, $\cos (t_{1,3}) = \sin (t_{1,3})  = 1/\sqrt{2}$.

During the second stage, qubit-1 interacts with qubit-2 while qubit-3 interacts with qubit-4. These two interactions may take place at the same time, in parallel. For each interaction, the time period is fixed such that the likelihood of excitation-transfer is equal to that of excitation-retention, i.e., $\cos(t_{1,2}) = \sin (t_{1,2})  = \cos (t_{3,4}) = \sin (t_{3,4}) =1/\sqrt{2}$. This yields the target state, with each qubit having the same amplitude ($1/2$) for carrying an excitation. 

For any larger value of $M$, we have $M$ stages. It can be easily seen that the total number of pairwise interactions is still $N-1$, the same as for the previous two protocols. However, all interactions within a stage may take place in parallel.  
Each of these interactions takes place over a time period given by $\cos (t) = 1/\sqrt{2}$, i.e., $t = \pi/4$.

If the setup is such that only one pairwise interaction can take place at a time, the total operating time would be $(N-1)\pi/4$, scaling linearly with $N$. If parallel pairwise interactions are possible, they significantly reduce operating time. We have $M$ distinct stages in the problem, each involving pairwise interactions over a time period of $\pi/4$. The lowest time is achieved if all interactions of a stage are performed simultaneously. This yields a lower bound for the operating time, $t_{lower~bound} = M \pi/4$. This quantity scales as $M \sim \log_2 N$.  

\begin{figure}
\includegraphics[width=3.3in]{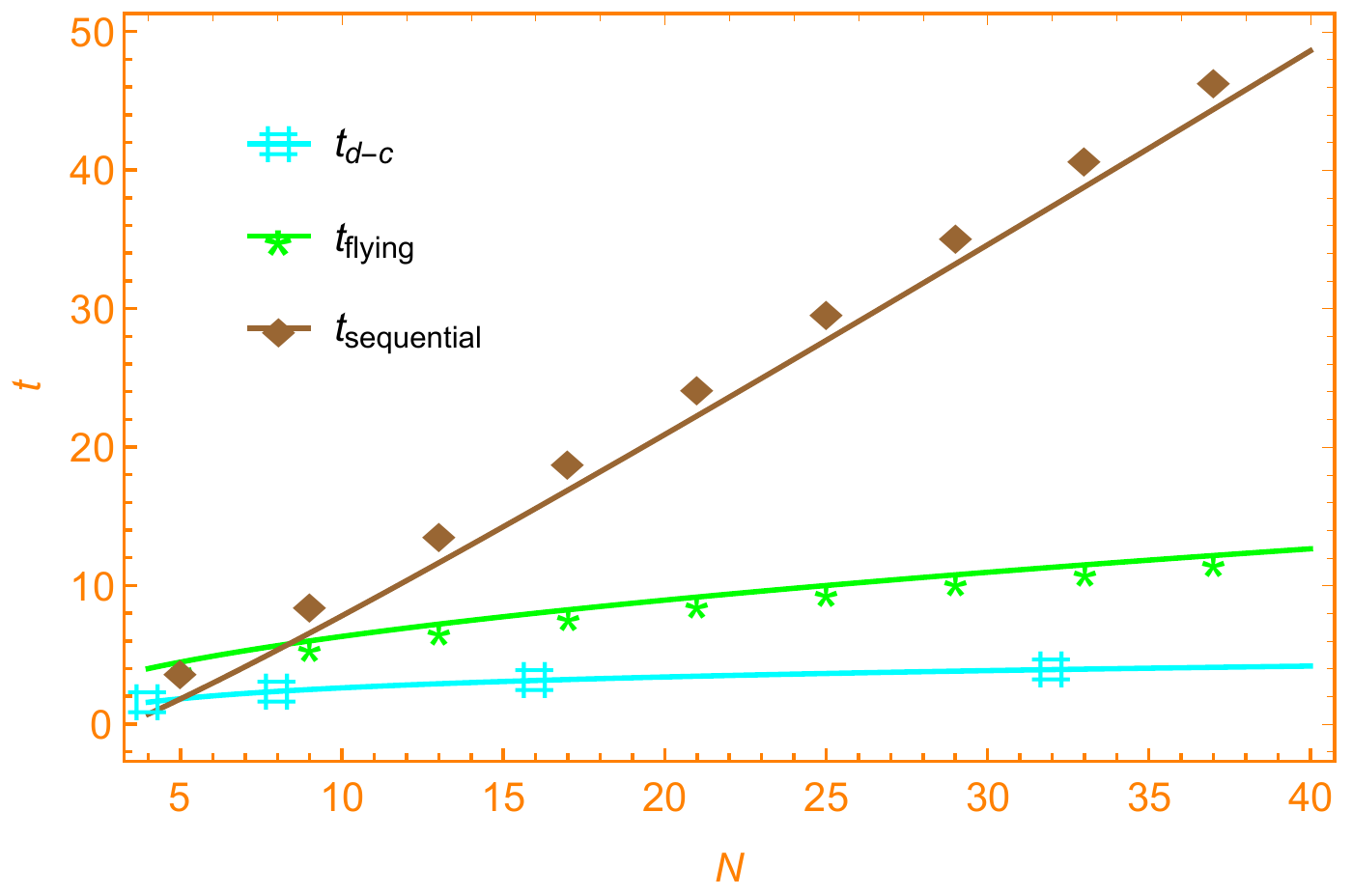}
\caption{The total interaction time $t$ vs. the number of qubits, $N$, for the three protocols discussed here. The lines show the approximate scaling form for large $N$ (see text). }
\label{fig.nvst}
\end{figure}

\section{Phase differences in the target state}
The target state, as defined in Eq.~\ref{eq.target}, has $N$ phases denoted as $\phi_j$'s. These phases cannot be independently controlled. With the exchange interactions of Eqs.~\ref{eq.U} and \ref{eq.Ulambda}, every excitation-transfer carries a phase of $3\pi/2$ (a factor of $-i$). 
As a result, in any of the three protocols, the final state corresponds to Eq.~\ref{eq.target} with disparate values of $\phi_j$'s. To illustrate this, we formally show that a W state (with all $\phi_j$'s being equal) cannot be synthesized using pairwise exchange interactions.

Our argument is based on two observations: (i) Pairwise exchange interactions are unitary operations and therefore, reversible. (ii) The W state is invariant under any permutation. As a result, it is unchanged (up to a global phase) by operators of the form Eq.~\ref{eq.U} or Eq.~\ref{eq.Ulambda}. Suppose the W state could be synthesized starting from a direct product state via pairwise interactions. It must be possible to reverse the process -- to start from a W state and to arrive at a direct product state with only pairwise interactions. However, this is not possible, as any operation of the form of Eqs.~\ref{eq.U} or \ref{eq.Ulambda} does not change the W state. We conclude that the W state cannot be produced within this approach. 

\section{Discussion}
We have discussed protocols to generate highly entangled generalized W-states starting from a direct product state. The entanglement is generated by pairwise interactions: at each step, two qubits are brought together and allowed to interact for a controlled period of time. This setup imposes strong constraints on the protocol and on the final state. For example, while it allows for an equal-weight superposition, the phase difference across components cannot be tuned. However, these constraints lead to highly-structured solutions. The time-intervals involved have a geometric interpretation, as angles of the spiral of Theodorus.

Our first protocol assumes a flying-qubit setup. This could be realized as a cavity-photon-mode that can be tuned into resonance with a series of static qubits\cite{Landig2019,Borjans2020}. In fact, this protocol has been used in Ref.~\onlinecite{Haffner2005} to create a generalized-W state with trapped ions. The second protocol involves sequential inter-qubit interactions. A similar protocol has been realized in Ref.~\onlinecite{Yadav2019} with semiconductor spin-qubits, but with the goal of transferring a qubit-state between the ends of a chain. The same setup can employ our sequential-protocol to generate a generalized W-state. Other studies have considered a qubit-chain with all interactions present concurrently. They have examined the transport of a qubit-state from one end of the chain to the other\cite{Ferron2022}.

We have discussed three protocols and compared them in terms of the time required.Fig.~\ref{fig.nvst} shows how the time required for each protocol scales with the number of qubits. In our analysis, we have only considered the total interaction time. An experimental realization will invariably have overheads associated with moving qubits, reconfiguring interaction circuits, etc. Future realization-specific studies could take these times into account. We have also ignored issues of adiabaticity, assuming that exchange interactions evolve the system with perfect fidelity. In practice, errors may arise from varying the interactions with time (e.g., see Ref.~\onlinecite{Sagesser_2020}). Despite these limitations, our study provides estimates that could inform protocol design. On a given quantum-computing platform, all the protocols considered here may not be accessible. For example, in ultracold atoms in optical lattices, qubits are arranged along a line. The divide-and-conquer algorithm may be ill-suited to such a setup as it requires qubits to be moved by large distances. Nevertheless, our results provide an estimate for the time involved.  

\acknowledgments
We thank Jean-S\'ebastien Bernier and G. Baskaran for insightful discussions. RG was supported by Discovery Grant 2022-05240 from the Natural Sciences and Engineering Research Council of Canada.
\bibliographystyle{apsrev4-1} 
\bibliography{Dicke_entanglement}
\end{document}